# *Nonvolatile plasmonics based on optically reprogrammable phase change materials*

Jacek Gosciniak

*Independent Researcher, 90-132 Lodz, Poland*
*Email: jeckug10@yahoo.com.sg*

**Abstract**
We propose here a new platform for a realization of novel nonvolatile optical switching devices that takes an advantage of high field confinement provided by plasmonics and multi-state programming capabilities of chalcogenide phase change materials. A high reduction in the overall energy consumption consists of a high field enhancement provided by plasmonic that allow to lower the switching energies and implementation of phase change materials that allow to operate under a zero-static power consumption. A combination of plasmonics and phase change materials provide additionally an essential improvement in terms of a switching time, attenuation contrast and possibility to perform a phase shift with the wide bandgap phase change materials.

In most of the all-optical switching photonic devices, a switching mechanism is realized optically through heating of phase change materials. Here, two stage heating process is proposed that is based on the absorption of light by phase change materials itself, and a heat transfer from the metal stripe under an absorption of light by a metal. Thus, compared to any other previously presented optical switches, even a wide bandgap phase change materials that show zero absorption of light can be implemented in the proposed structure.

The proposed plasmonic waveguide arrangement is extremely sensitive to any changes of the phase change material properties, thus, even a minor change of temperature provides an essential change in the transmitted light.

**Introduction**
With a growing demand for data processing, the search for new computing technologies arise very quickly in a last 10 years. Traditional computing is based on the von Neumann architecture that consists of a central processing unit (CPU) which carriers out the computations and a separated memory that stores data and instructions [1]. They suffer however from separation between processing and memory what limits the performance of the overall computing system. Neuromorphic computing [2] and in-memory computing [3, 4] are two such approaches that bring processing and memory operations together and enable parallel processing.

Neuromorphic computing architecture try to mimic the working principles of the human brain through integrated circuits [2-4]. Thus, similarly to human brain, neuromorphic processors aim to work in a highly parallel way and process data directly in memory [3, 4].

There are two main directions in development of a neuromorphic hardware that are based on electronics [5-7] and photonics [8-11]. Electronics offer a high-degree physical interconnections through a dense mesh of wires in a form of crossbar arrays with memristors [4, 6, 7, 12-14] integrated at the junctions of crossbar arrays [5-7]. They suffer however from a low bandwidth and low speed [15]. In comparison, photonics offers a huge bandwidth and high speed but suffer from large footprint and low-level photonic integration [16-19]. Apart from it, both electronics and photonics can operate only either in electronic or photonic domains. Thus, it will be highly beneficial to develop a new hardware that is based on plasmonics as it can serve as a bridge between the electronics and photonics domains [20]. Furthermore, it will be highly beneficial to take an advantage of crossbar arrays with memory cells as it offers extremally low energy per bit that scales with a feature size [21]. Apart from it, to overcome some of the limitations and get closer in the way of operation to the human brain, the nonvolatile functionalities are essential [8, 9, 22]. A term nonvolatile means that no static energy or holding power is required to retain any of the states once it is set [23, 24]. Nonvolatility can be only realized through materials, thus, nonvolatile materials are under a deep interest [23].

Phase-change materials (PCMs) present a very pronounced contrast in the electrical conductance and the complex refractive index between their amorphous and crystalline phases that can be induced via electrical or optical heating [25-33], what ensures a dual electrical-optical functionality [34-36]. Furthermore, those changes are reversible, fast and can be performed over many cycles [29, 33].

Here, we propose the plasmonic waveguides with the PCM placed either in the ridge or buffer layer depending on the requirements and fabrication possibilities. Such devices enable to control the optical amplitude and/or phase of the transmitted signal by controlling a phase of the PCM under an interaction with a light. In a case of plasmonic, the interaction of light with the PCM is highly enhanced, thus, a high reduction in power and time can be expected [20, 34, 35]. The state of the device with the PCM cell can be optically controlled, "read", as a transmission of light depends on the phase of the PCM. Furthermore, the PCM cell can also be controlled optically by sending appropriate "write", amorphization, and "erase", crystallization, pulses down the waveguide [26, 28-42]. Besides, as the switching process shows a threshold behavior, it can be used as a nonlinear element inside a photonic circuit [26].

Most of the available phase change photonic devices suffer from high switching energies and long switching times in the tens of hundreds of nanoseconds [44]. In comparison, a proposed here phase change plasmonic devices can reduces switching energy requirements and increase switching speed through more efficient heating of the PCM and stronger dependance of the PCM state on the transmission of light. Furthermore, the change in a transmission of light is achieved not only by a manipulation of the imaginary part of the PCM but through a change in a real part of the PCM as well [35, 36]. Besides, the interaction of light with the PCM is highly magnified as PCM is arranged in the plasmonic waveguide.

**Motivation**

To justify our interest in a development of new plasmonic platform for on-chip optical switching, we perform a short comparison between photonics and plasmonics in terms of desired properties for this type of switches.

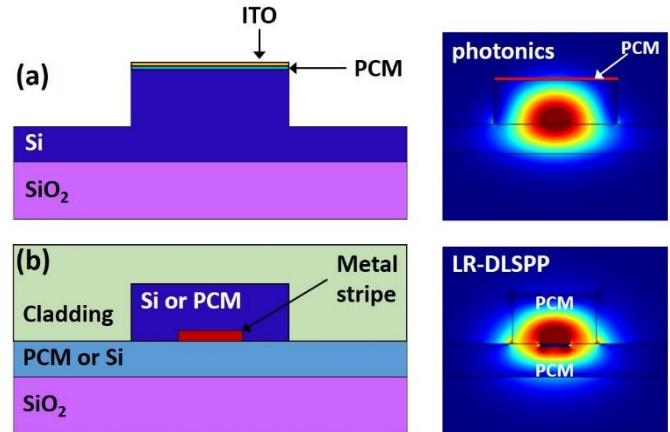

**Figure 1.** The cross section of the all-optical (a) photonic and (b) plasmonic memory elements with a corresponding simulated eigenmode profiles of the (a) fundamental TE mode inside Si waveguide with PCM on top of the waveguide and (b) fundamental TM mode inside LR-DLSPP waveguide with PCM placed either in a buffer layer or a ridge.

In a photonic waveguide, a thin layer of PCM (10 nm GST) is deposited on top of the Si or SiN waveguide and covered with indium tin oxide (ITO) (10 nm thick) to prevent oxidation of the PCM and force a TE mode to PCM to enhance absorption (Fig. 1a) [26, 42]. As observed from Fig. 1a, the PCM is placed far away from the electric field maximum of the propagating mode, thus, the interaction of light with the PCM is weak.

In comparison, for a long-range dielectric-loaded surface plasmon polariton (LR-DLSPP) waveguide, the PCM can be placed either in a buffer layer or ridge and in direct contact with a metal stripe, *i.e.*, in the electric field maximum of the propagating plasmonic mode (Fig. 1b). Furthermore, as the electric field is highly enhanced in the plasmonic waveguide, the electric field of the propagating LR-DLSPP mode (Fig. 1b) that interact with the PCM is much stronger compared to the photonic waveguide (Fig. 1a).

Furthermore, as it has been mentioned previously [35], the optical switching in a proposed waveguide can be highly enhanced through a heat transfer from a metal stripe under an absorption of light to the PCM that is in a direct contact with a metal stripe [43, 45]. Thus, two stage heating process of PCM can take place in a proposed LR-DLSPP arrangement that is related with (a) a direct absorption of light by PCM, and (b) a heat transfer from a metal stripe to the PCM under an absorption of light by metal stripe. Thus, even a wide bandgap PCMs, such as for example SbSe that have zero absorption in the near IR [29, 46], can be all-optically switched taking an advantage of plasmonics and presence of the internal metal stripe that is a part of the LR-DLSPP waveguide. The optical power absorbed by the metal stripe is dissipated into any materials that are in contact with a metal stripe and an amount of heat dissipated into PCM depends on thermal resistance and capacity of the surroundings materials [35, 43]. The dissipated power by metal stripe depends on the surface plasmon polariton (SPP) attenuation coefficient and the length of the active region. Thus, the higher attenuation coefficient, the higher increase of the metal stripe temperature.

**Photonic-plasmonic integration**
A very essential part of each photonic system it to provide an efficient amount of light to the active area of the system. Even very efficient photodetectors or modulators can suffer from the overall efficiency under low coupling efficiency of light to such devices. Thus, for example, a waveguide-integrated photodetectors suffer from a low external efficiency which translates on the photodetector responsivity. In many situations a low coupling efficiency means that higher power is needed to provide an efficient amount of light to a system which, however, increases the overall energy budget of the system. This point is of particular importance in relation to plasmonics that usually requires special attention as it is usually a low compatibility between photonics and plasmonic systems. Thus, a coupling efficiency from photonic to plasmonic waveguides usually is below 50 % what highly reduces the energy budget of the system and put a lot of restrictions on the system [20, 34, 47, 48].

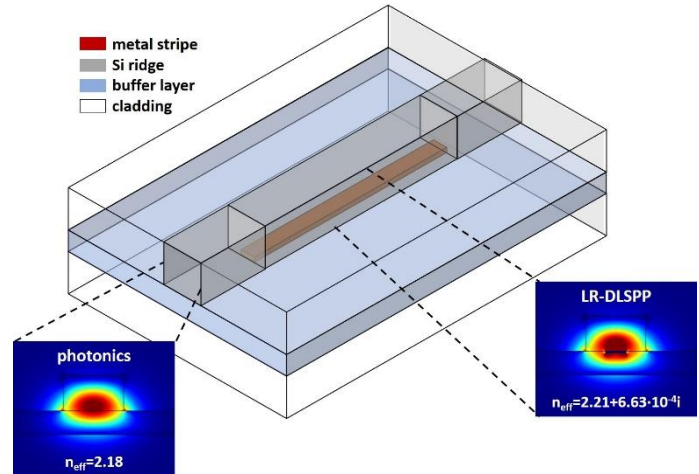

**Figure 2.** Coupling schema between photonic and LR-DLSPP waveguides with a corresponding Poynting vectors. The mode effective indices were calculated for Si ridge dimensions $w = 300$ nm and $h = 120$ nm, and buffer layer thickness $h = 120$ nm.

Here we present a very efficient coupling arrangement from photonic waveguide to a plasmonic active component that do not require any special transfer system or additional fabrication steps (Fig. 2). The proposed LR-DLSPP waveguide is similar to the photonic rib waveguide with a thin metal stripe placed between a buffer (rib) and a ridge. The maximum electric field of the photonic waveguide coincides highly with the maximum electric field of the plasmonic waveguide, thus, an extremally high modes matching conditions are fulfilled [49, 50].
In a proposed all-optical switching realized with LR-DLSPP waveguide, a light from a photonic waveguide is very efficiently delivered to the plasmonic all-optical switch with the coupling efficiency exceeding 95 % [49, 51]. In consequence, the power losses are kept at the minimum and most of the power can be involved in a switching process.

**Optical switching**

Mechanism of switching in the PCM is realized through a phase transition between its amorphous (high resistance and high transmission) and crystalline (low resistance and low transmission) phases under a heating of PCM [25, 31]. It can be performed by Joule heating using external heaters [27, 52-54], electrical "memory" switching by connecting the PCMs into a circuit [34], or optical switching using the optical pulses [26, 39-42]. In on-chip photonic nonvolatile PCM switches, the heat-induced refractive index change of PCM entail a change in transmitted light by modulating amplitude or optical phase [35].

In most of the optical switching devices the switching process is realized through an absorption of light by PCM that show reasonably high an imaginary part of its refractive index. Under an absorption, the light is converted into heat that cause a phase transition of PCM. Thus, most of the devices that are based on the optical switching mechanism is limited to a low bandgap PCMs. In consequence, a wide bandgap PCMs such as SbS and SbSe are not practical for such applications.

In this point it is very valuable to notice, that while only some PCMs exhibit an imaginary part of the refractive index or show a sufficient change in an imaginary part under a phase transition, all of them show significant change in a real part of the refractive index [25, 26, 28, 29]. However, up to now, this obvious observation has remained without consequences with respect to devices design. In this paper we show that complex PCMs refractive index can be involved in a heat generation and even wide bandgap PCMs can be considered for the optical switching. In the past, the presence of metal and associated with it losses were considered as a limiting factor for implementation of plasmonics in real devices. However, in the last 10 years became clear that it can provide some benefits such as, for example, in the generation of hot electrons [55] or in the catalysis [56]. In this paper we show that it can be used for a heat generation and enhancement of the temperature of the PCMs.

Under an absorption of light by a metal stripe, the stripe is heated and then, the heat is dissipated to any materials that are in contact with a metal stripe. The amount of generated heat depends on the absorbed power by metal stripe. When a balance between mode effective indices on both sides of metal stripe is disturbed, the absorption losses in metal arise and more heat is transferred to the surrounding materials – the higher unbalance, the higher heat transferred to PCM. Thus, two stage heating process can be involved in a phase transition of any PCMs in a proposed LR-DLSPP arrangement: (a) direct absorption of light by PCMs as result of an imaginary part of the refractive index and, (b) indirect heating of PCMs as result of a heat transfer from a metal stripe that is purely related with the real part of the PCMs refractive index.

In a proposed plasmonic arrangement, the PCM is deposited either in a buffer layer or in a ridge and is part of the plasmonic waveguide (Fig. 1b). Optical pulses, *i.e.*, "input" pulses, send down the photonics waveguide couple efficiently to the plasmonic waveguide and, consequently (thus), to the PCM (Fig. 2). Under a heat transfer to the PCM, the PCM is switched between its amorphous and crystalline states, or one of many intermediate levels of crystallinity lying between these states. Thus, the optical transmission of the waveguide is programmed by the "input" pulses since the transmission depends on the refractive index of PCMs that is highly phase-state dependent. Then, the "probe" pulse can be used in the readout process to determine the information stored in the PCM cell by the "input" pulses (Fig. 3) [40].

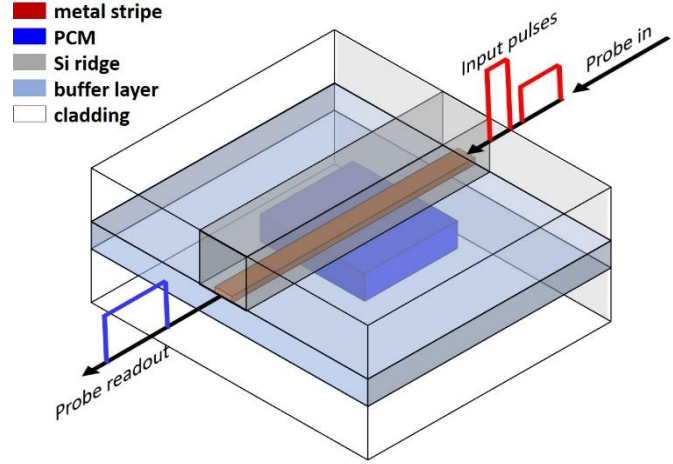

**Figure 3.** Schematic of the proposed plasmonic phase-change memory device with the operation principles.

The "input" pulses can be used in "write" or "erase" modes, where "write" pulses are used for amorphization of PCM and "erase" pulses for crystallization. Such multi-state PCM arrangement can be incorporated into novel plasmonic crossbar arrays [57] to deliver ultra-fast matrix vector multipliers and to realize synaptic and neuronal "mimics" and small-scale neuromorphic processors.

To analyze the change in a transmission of light through the plasmonic waveguide under a change of the PCM refractive index placed either in a ridge or buffer layer, detailed calculations were performed for different ridge dimensions and buffer layer thickness (Fig. 4 and 5).

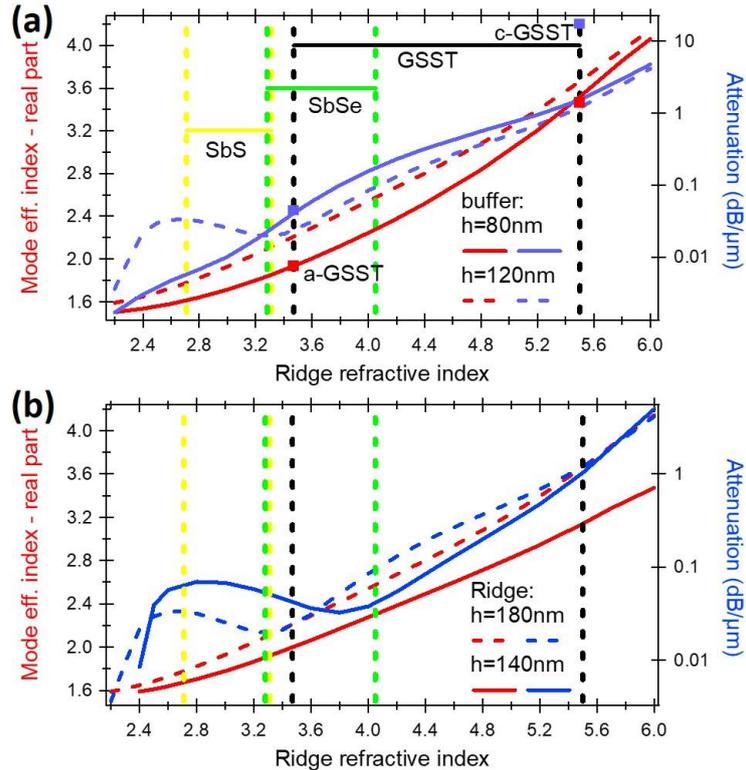

**Figure 4.** The mode effective index and attenuation as a function of PCM ridge refractive index for (a) different Si buffer layer thickness and (b) different PCM ridge height. (a) The PCM ridge height and width were kept at $h$ = 180 nm and $w$ = 300 nm, respectively. (b) The ridge width was kept at $w$ = 300 nm while the buffer layer thickness at $h$ = 120 nm.

Here, the color of the vertical dashed lines (yellow, green, and black) represents the working range of a given PCM, *i.e.*, a refractive index of an amorphous and crystalline state of PCMs, as showed at Fig. 4a. Besides, the curves stand for calculations performed under assumption of no loss PCMs while the markers stand for calculations performed for complex GSST. It means that for all curves in Fig. 4 and 5, a real part of the refractive index was swept from $n = 2.2$ to $n = 6.0$, while an imaginary part was taken as zero. There were two reasons for that – first, SbS and SbSe shows zero absorption under wide wavelength range for both an amorphous and crystalline states, and second, it is hard to evaluate the intermediate values of the complex refractive index of GSST.

For GSST in the amorphous phase, the imaginary part of the effective index is significantly lower ($n_{im} = 0.0002$) compared to the crystalline state ($n_{im} = 0.42$) (Table 1), leading to a large absorption contrast between both modes: $\alpha_{a\text{-}GSST} = 0.044$ dB/μm and $\alpha_{c\text{-}GSST} = 17.2$ dB/μm (Fig. 4a). It means over 390 times increases in an attenuation of the LR-DLSPP mode under a phase transition of GSST. When compare with a photonic waveguide realized with the GST on top of Si waveguide, the attenuation for amorphous and crystalline states of GST were calculated at $\alpha_{a\text{-}GST} = 0.07$ dB/μm and $\alpha_{c\text{-}GST} = 2.96$ dB/μm, respectively [26]. It means around 42 times increases in an attenuation of the photonic mode under a phase transition of GST. Furthermore, the attenuation of LR-DLSPP waveguide with the GSST placed in a ridge and at a crystalline state is over 5 times higher compared to the photonic waveguide realized with the GST at a crystalline state, despite even 2.5 times higher an imaginary part of the refractive index of GST ($n_{im} = 1.089$) compared to GSST ($n_{im} = 0.42$).

Under an absorption of light by metal stripe and efficient heat transfer from a metal stripe to the PCM, even a zero-loss PCMs, such as SbS and SbSe, are able to show a change in attenuation under a transition of PCM. As showed in Fig. 4, the attenuation of SbS changes from $\alpha_{a\text{-}SbS} = 0.0334$ dB/μm for SbS at an amorphous state to $\alpha_{c\text{-}SbS} = 0.0196$ dB/μm for SbS at a crystalline state and for a Si buffer layer thickness of 120 nm. Thus, compared to GSST, the attenuation of SbS at the amorphous state is higher than in a crystalline state. However, for a Si buffer layer thickness of 80 nm, the attenuation of SbS at a crystalline state is higher than attenuation at amorphous state ($\alpha_{a\text{-}SbS} = 0.0057$ dB/μm and $\alpha_{c\text{-}SbS} = 0.024$ dB/μm) (Fig. 4a). In terms of SbSe, the attenuation of SbSe for a LR-DLSPP waveguide with Si buffer layer thickness of 120 nm increases from $\alpha_{a\text{-}SbSe} = 0.0196$ dB/μm for SbSe in an amorphous state to $\alpha_{c\text{-}SbS} = 0.09$ dB/μm for SbSe in a crystalline state. However, for SbSe ridge height of 140 nm, the attenuation of LR-DLSPP waveguide with the SbSe at an amorphous state was calculated at $\alpha_{a\text{-}SbSe} = 0.05$ dB/μm while at a crystalline state at $\alpha_{c\text{-}SbS} = 0.048$ dB/μm (Fig. 4b). Simultaneously, the mode effective index change was calculated at $\Delta n_{eff} = 0.4$. Thus, an absorption of light by LR-DLSPP waveguide with the SbSe placed in a ridge may lead to a phase shift of light rather than an attenuation under a light transmission through a LR-DLSPP waveguide. Consequently, to provide a full π phase shift only $L = 1.94$ μm long LR-DLSPP waveguide is required while an insertion loss (IL) is kept below 0.1 dB. This represents a tremendous turnaround in the realization of all-optical switching that until now has only been possible to attenuate an optical signal.

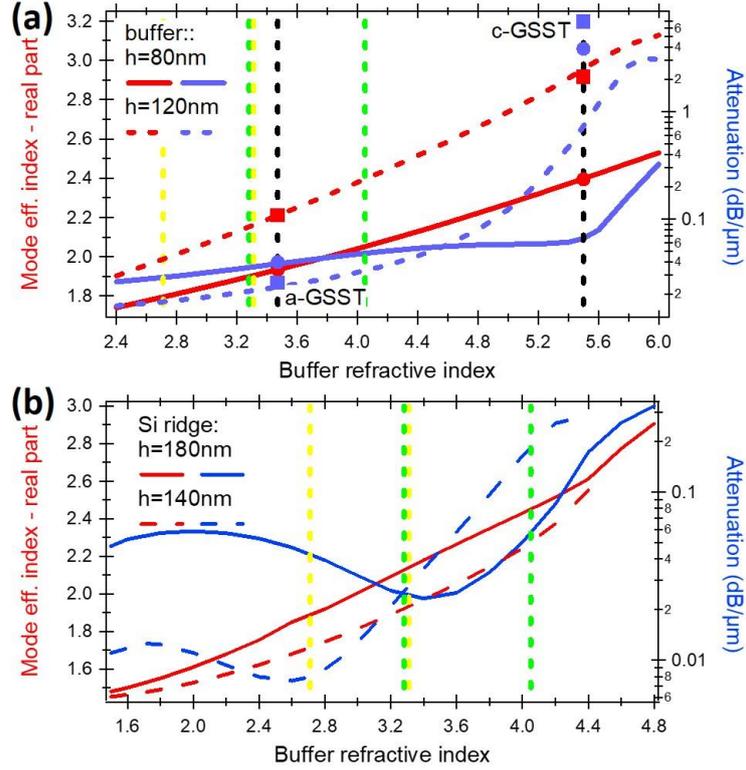

**Figure 5.** The mode effective index and attenuation as a function of PCM buffer layer refractive index for (a) different thickness of PCM and (b) different thickness of Si ridge. (a) The Si ridge height and width were kept at $h = 180$ nm and $w = 300$ nm, respectively. (b) The Si width was kept at $w = 300$ nm while the PCM buffer thickness at $h = 120$ nm.

In a second scenario, the PCM was placed in the buffer layer while the Si was implemented in a ridge. In Fig. 5a, the width of the heated buffer layer area was limited in size to the width of the ridge ($w = 300$ nm) while in Fig. 5b the width increased to $w = 800$ nm. This is due to the fact it is very difficult to estimate in this case the size of the area that has an essential influence on the change of the mode effective index and attenuation. Based on our previous experience [58-60], we can assume that only a small part of the PCM that is in direct contact with the metal stripe has an essential influence on the characteristic of a transmitted light, *i.e.*, the area of PCM placed in the electric field maximum of the propagating mode.

The attenuation of LR-DLSPP waveguide with a GSST placed in a buffer layer was calculated at $\alpha_{a\text{-}GSST} = 0.025$ dB/μm for a GSST in an amorphous state (a-GSST) while at a crystalline state (c-GSST) it was calculated at $\alpha_{c\text{-}GSST} = 6.915$ dB/μm (Fig. 5a). As in a previous scenario with a GSST placed in a ridge, it is again much higher compared to the photonic structure with the GST placed on top of the Si or SiN waveguide [26].

As seen from Fig. 5, the level of attenuation for different PCMs can be mostly exactly defined through a design by a proper choice of the ridge dimensions and thickness of the PCM buffer layer.

From a point of view of implementation this technology it will be beneficial to design such the all-optical memory device with the PCM in a crystalline state at an initial state as it is characterized by higher absorption of the LR-DLSPP waveguide. Thus, the absorption losses in metal arises and heat generation increases that is then transferred to the PCM. In consequence, the power needed to provide a phase transition of PCM decreases and device can operate under lower power conditions.

**Heat transfer**

Under an absorption of light by a metal stripe (Fig. 6), the optical energy is transferred into heat. The heat from metal stripe dissipates to any materials that are in contact with the metal through conductive heat transfer [43].

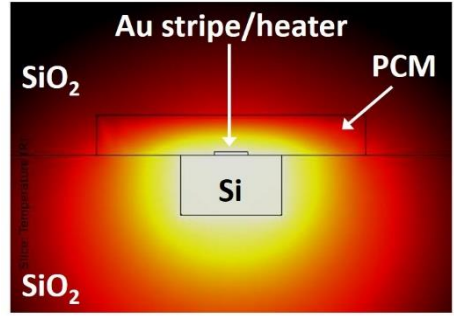

**Figure 6.** Heat map produced by optical switching under an absorption of light by metal stripe with the PCM placed in a buffer layer. The Si ridge dimensions were kept at $w$ = 300 nm and $h$ = 180 nm while the PCM-based buffer layer at $w$ = 800 nm and $h$ = 120 nm.

The amount of heat transfer to the area of interest (ridge or buffer layer) depends upon the thermal conductivity coefficients of any materials that are in contact with metal stripe, contact area and thickness of the materials. Thus, to ensure an efficient heat transfer to the PCM, the materials that are in contact with metal stripe should possess low thermal conductivity coefficient [43]. Furthermore, direct contact of PCM with the electrodes in the proposed modulator allows lowering the threshold voltage for delivering the right amount of heat for inducing a phase transition in the PCM.

**Table 1.** Thermal properties of the materials forming the device. Here, $T_c$ is a crystallization temperature and $T_m$ is a melting temperature of PCM. The refractive indices were provided for wavelength of 1550 nm.

| Material | Refractive index | Thermal cond. coeff. (W/m·K) | Heat capacity (J/g·K) | Density (g/cm$^3$) | $T_c$ (°C) | $T_m$ (°C) |
|---|---|---|---|---|---|---|
| Air | 1 | 0.026 | 1.005 | - | | |
| SiO$_2$ | 1.45 | 1.38 | 0.746 | 2.19 | | |
| Si | 3.47 | 148 | 0.72 | 2.32 | | |
| Si$_3$N$_4$ | 1.996 | 20 | 0.7 | 3.1 | | |
| Au | 0.5958+10.92$i$ | 318 | 0.130 | 19.32 | | |
| a-GST | 3.80+0.025$i$ | 0.19 | 0.213 | 5.87 | 160 | 630 |
| c-GST | 6.63+1.089$i$ | 0.57 | 0.199 | 6.27 | | |
| a-GSST | 3.47+0.0002$i$ | 0.17 | 0.212 | 6.0 | 523 | 900 |
| c-GSST | 5.50+0.42$i$ | 0.43 | 0.212 | 6.15-6.3 | | |
| a-SbS | 2.71+0$i$ | - | - | - | 270 | 550 |
| c-SbS | 3.31+0$i$ | 1.16-1.2 | 0.353 | 4.6 | | |
| a-SbSe | 3.28+0$i$ | 0.36-1.9 | 0.507 | 5.81 | 180 | 620 |
| c-SbSe | 4.05+0$i$ | - | 0.574 | - | | |

To evaluate a heat transfer to the PCM the thermal resistance and thermal capacitance should be considered. The thermal resistance of an object is defined as $R_{th} = L/(\kappa \cdot A)$ and it describes the temperature difference that will cause the heat power of 1 Watt to flow between the object and its surroundings. In comparison, the thermal capacitance of an object, $C_{th} = C_p \cdot \rho \cdot V$, describes the energy required to change its temperature by 1 K, if no heat is exchanged with its surroundings [35, 43]. Here, $L$ is the length of substrate along the heat transfer direction, $A$ is the cross-section area of the substrate, $V$ is the heated volume of the material, $\kappa$ is the thermal conductivity of the material, $C_p$ is the specific heat and $\rho$ is the mass density. Thus, the lower thermal capacitance the higher temperature rises for a given amount of heat delivered to the material. Furthermore, materials with lower thermal conductivity coefficient are characterized by higher thermal resistance, so lower electrical powers are required to increase a material temperature as the heat loss is reduced.

As it has been mentioned previously [35], even extremally small area of PCM that is in direct contact with a metal stripe is able to provide a significant change in a mode attenuation during a heat transfer from a metal stripe under an absorption of light (Fig. 7).

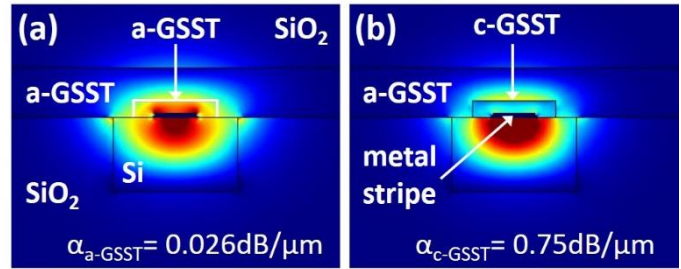

**Figure 7.** Simulated eigenmode profiles of the fundamental TM modes showing the field distribution inside a plasmonic waveguide with corresponding mode attenuations for an assumption only part of GSST in (a) an amorphous and (b) crystalline phases.

Here, the heated area of GSST was limited in size to only 30 nm above a metal stripe and 50 nm on both sides of metal stripe. Despite, over 30 times increases in a plasmonic mode attenuation was calculated that rises from $\alpha_{a\text{-}GSST}$ = 0.026 dB/μm to $\alpha_{c\text{-}GSST}$ = 0.75 dB/μm under heating of GSST and, consequently, its phase transition (Fig. 7). A significant increase of an attenuation, even though a small area of PCM, is due to the presence of the PCM in direct contact with a metal stripe, *i.e.*, in the electric field maximum of the propagating mode what greatly enhances the sensitivity. Furthermore, the electric field of the plasmonic waveguide is much stronger than in its photonic counterpart what provides additional benefits as the heat generation arises from a stronger light-PCM interaction. Additionally, it allows to minimize the switching voltage and current of PCM devices and, consequently, lower power consumption [33].

All calculations presented here were performed for a gold metal stripe width $w$ = 100 nm and thickness $h$ = 10 nm. However, using thicker or wider metal stripes can be beneficial for this type of devices as the absorption losses in metal arise with an increase the metal stripe thickness and width. Furthermore, we focused in this paper on gold (Au) but it can be replaced by other "plasmonic" metals such as, for example, a CMOS-compatible titanium nitride (TiN) [61-63].

**Conclusion**
In conclusion, we have proposed a new class of optically driven plasmonic nonvolatile switches realized with the phase change materials that operate under a huge attenuation contrast and offering a phase shift. Such optical switches operate under a zero-static power consumption while being very compact. A huge improvement in terms the overall performances can be observed when compared with photonic devices. In photonic devices that operate under optical switching mechanism, the PCM is placed on top of the waveguide, *i.e.*, far away from the electric field maximum of the propagating mode. In comparison, the field enhancement of the plasmonic waveguides is much stronger compared to its photonic counterparts, and furthermore, the PCM is placed directly in the electric field maximum what highly enhances the interaction of light with the PCM. Thus, even a small area of PCM close to the metal stripe can provide an essential influence on the attenuation as the proposed plasmonic waveguide is extremely sensitive to any changes of the mode effective indices on both side of the metal stripe.

Additionally, the PCM heating mechanism is not limited only to the absorption of light by PCM, as in the case of photonics, but is highly enhanced through a heat transfer from the metal stripe under an absorption of light by a metal. Thus, even wide bandgap PCMs that show zero absorption can be considered for the optical switching. In addition to attenuation, the proposed plasmonic waveguide offers to provide a phase shift when operate in the optical switching mode, while the attenuation is kept constant at low level.

**Author information**
**Affiliations**
Independent Researcher, 90-132 Lodz, Poland
Jacek Gosciniak
**Contributions**
J.G. conceived the idea, performed all calculations and FEM and FDTD simulations and wrote the article.
**Corresponding author**
Correspondence to Jacek Gosciniak (jeckug10@yahoo.com.sg)